\documentclass{article}
\usepackage{spconf,amsmath,amssymb,graphicx,booktabs,float,url,multirow,fancyhdr}
\usepackage[hidelinks]{hyperref}

\def\L{{\cal L}}
\newcommand{\e}{\mathbf{e}}
\newcommand{\m}{\mathbf{m}}
\newcommand{\z}{\mathbf{z}}

\floatstyle{plaintop}
\newfloat{algorithm}{tbp}{loa}
\floatname{algorithm}{Algorithm}
\newcounter{algline}
\newenvironment{pseudocode}
  {\setcounter{algline}{0}\small\setlength{\parskip}{0pt}\setlength{\parindent}{0pt}}
  {}
\newcommand{\alghead}[1]{{\small\raggedright\par\noindent
  \hangindent=1.2em\hangafter=1 #1\par}\vspace{1pt}\hrule\vspace{2pt}}
\newcommand{\algline}[2][0em]{%
  \par\noindent\refstepcounter{algline}%
  \hangindent=\dimexpr 2.1em+#1\relax\hangafter=1
  \makebox[1.6em][r]{\small\thealgline:}\hspace{0.5em}\hspace*{#1}#2\par}
\newcommand{\algcom}[1]{\hfill\textit{\small$\triangleright$ #1}}
\newcommand{\kw}[1]{\textbf{#1}}

\title{FASTDIAR: FrAme-level speaker encoder for STreaming DIARization}

\twoauthors
 {Nikita Torgashov\textsuperscript{*}\thanks{\textsuperscript{*}Work done during internship at Microsoft.}}
	{Department of Speech, Music and Hearing\\
	KTH Royal Institute of Technology\\
	Stockholm, Sweden}
 {Okan Köpüklü}
	{Microsoft\\
	Munich, Germany}

\fancypagestyle{postprint}{
  \fancyhf{}
  
  \fancyfoot[C]{
    \begin{minipage}{\textwidth}
    \begin{footnotesize}\begin{center}
    \textcopyright{} 2026 IEEE. Personal use of this material is permitted. Permission from IEEE must be obtained for all other uses, in any current or future media, including reprinting/republishing this material for advertising or promotional purposes, creating new collective works, for resale or redistribution to servers or lists, or reuse of any copyrighted component of this work in other works.
    \end{center}\end{footnotesize}
    \end{minipage}
  }
}

\begin{document}

\makeatletter
\patchcmd{\thebibliography}
  {\advance\leftmargin\labelsep}
  {\setlength{\itemsep}{1.5pt plus 0.3pt}
   \setlength{\parsep}{1pt}
   \setlength{\parskip}{1pt}
   \advance\leftmargin\labelsep}
  {}
  {}
\makeatother

\maketitle

\begin{abstract}
Real-time conversational agents require speaker diarization that streams and runs on
a CPU. Most systems apply an utterance-level speaker encoder to short, heavily
overlapping chunks, which wastes computation and leaves the model optimized for
the wrong task. We instead turn a state-of-the-art speaker recognition
architecture into a causal frame-level encoder that reads the stream once and
emits one embedding every 80~ms from a bounded two-second window of past audio,
and pair it with online clustering that gates every update on the
self-similarity of the stream. Trained only by distillation from an
utterance-level teacher on simulated and out-of-domain mixtures, and evaluated
with one fixed set of hyperparameters, the system is the most accurate streaming
diarizer on low-overlap benchmarks at sub-second latency, degrades far less than
cache-based systems as the number of speakers grows, and runs five times faster
than real time on a single CPU thread.
\end{abstract}


\section{Introduction}
\label{sec:introduction}

Speaker diarization answers who speaks when and is a core component of spoken
conversational systems. Personalized automatic speech recognition (ASR),
voice activity detection (VAD), and speech
enhancement all depend on knowing which speaker is active, and a live agent must decide
while the conversation unfolds, so low-latency streaming is a key requirement.
GPUs are often unavailable, either because the system runs on a low-power device
or because the GPU is busy with ASR, a large language model, or speech synthesis,
which makes CPU diarization desirable.

Most diarization systems are built around a speaker encoder
\cite{desplanques2020ecapa,yakovlev2024redimnet}, applied off the shelf to short, heavily overlapping
chunks. That is convenient but costly: consecutive windows re-encode every sample
many times over, and verification training assumes one speaker per utterance, so
the encoder is never asked to forget a speaker or localize a change.

Offline diarization is dominated by cascaded pipelines combining local
segmentation with clustering, such as Pyannote \cite{Bredin23} and DiariZen
\cite{han2025leveraging}, while end-to-end models predict activities directly
with permutation-invariant training \cite{fujita2019eend}, attractors
\cite{horiguchi2020eendeda} or arrival-time ordering \cite{park2024sortformer}.
All of them need the whole recording before emitting labels, so none can stream.

Existing streaming systems show recurring trade-offs. Streaming Sortformer
\cite{medennikov25_interspeech} keeps an arrival-order speaker cache and is
accurate, but it is capped at four speakers and its 117.7M parameters are too
slow on CPU. The speaker-tracing buffer \cite{xue2021stb} re-processes buffered frames at
every step, BW-EDA-EEND \cite{han2021bweda} is linear in time but degrades beyond
two speakers, and EEND-GLA \cite{horiguchi2023gla} lifts the speaker-count limit
without reporting a CPU budget. On the cascaded
side, diart \cite{coria2021diart} attains low latency on CPU with incremental
clustering over local segmentation, but is less accurate.

The encoder in front of the clustering usually stays segment-level. UIS-RNN
\cite{zhang2019uisrnn} makes clustering supervised but reads d-vectors from
windows with 50\% overlap, and Turn-to-Diarize \cite{xia2022turn} ties output
resolution to a turn detector. Two works reach frame rate:
\cite{cordlandwehr2023frame} distils an utterance-level teacher into a student
that pools over a centred window, and LS-EEND \cite{liang2025lseend,
liang2024fseend} learns a causal frame-wise encoder jointly with its attractor
decoder. Neither gives a streaming encoder with a bounded past: the first reads
the future, the second carries the whole recording in its recurrent state.

In this work we present FASTDIAR (FrAme-level speaker encoder for STreaming
DIARization). Our contributions are (i) a causal frame-level speaker encoder that needs no
chunking, no temporal pooling and no recomputation, and summarizes a two-second past window, (ii) an extension of incremental online clustering where an embedding-confidence gate decides which frames may shape a speaker
centroid, and (iii) FASTDIAR -- the most accurate streaming diarizer on
low-overlap benchmarks at sub-second latency, far more robust than cache-based
systems as the speaker count grows, and five times faster than real time on one CPU thread with no real diarization corpus
in training. The encoder is a drop-in front end whose embeddings can feed other
speaker-aware models. Code and models are available at \mbox{\url{https://github.com/microsoft/fastdiar}.}

\section{Method}
\label{sec:method}

\subsection{Streaming speaker encoder}
\label{subsec:encoder}

We take ReDimNet2 \cite{yakovlev2026redimnet2} as the baseline, a leading
speaker recognition architecture with a low FLOPs budget. It
alternates 2D convolutional blocks over a channel-frequency-time feature map
with 1D blocks that operate on the same map after reshaping, and it produces a
single utterance-level embedding through temporal pooling. We address three properties of
this design that conflict with streaming diarization.

\noindent\textbf{(1) Causal convolutions.} We pad every 2D convolution only on
the left of the time axis, so the output at frame $t$ depends on frames up to
$t$. This removes access to future context.

\noindent\textbf{(2) Fully connected 1D blocks.} The 1D attention blocks share
temporal information across the whole utterance, giving an unbounded receptive
field. That helps verification and hurts diarization, where the encoder must
forget the current speaker and follow a new one. We replace them with fully
connected blocks, which mix frequency bins and channels without extending the
temporal receptive field.

\noindent\textbf{(3) Frame-level embeddings.} Temporal pooling aggregates
statistics over the whole utterance and makes the past receptive field
unbounded. In streaming diarization each embedding must
instead rely on a limited number of past frames, so speaker changes are handled
promptly and cluster boundaries stay sharp. We remove the pooling layer and project the
encoder output directly to a $d = 192$ dimensional embedding at every step.
Raising the time stride of the final block from 1 to 2 lowers the FLOPs count and
makes each embedding summarize 80~ms of audio rather than 40~ms.

The resulting encoder has a receptive field of two seconds of past audio. It
reads the stream once and emits one $\ell_2$-normalized embedding $\e_t$ every
$\tau = 80$~ms, with no chunking and no recomputation of already seen samples.
Dropping pooling bounds our past and lets an embedding forget a speaker, at a
verification cost quantified in Section~\ref{sec:results}.

\begin{table*}[t]
\centering
\caption{Diarization results (DER, \%). Bold marks the lowest DER in each column within an overlap group.}
\vspace{0.1em}
\label{tab:diarization}
\small
\renewcommand{\arraystretch}{0.9}
\setlength{\tabcolsep}{3pt}
% tight gaps inside each speaker-count group, normal gaps between datasets
\newcommand{\spkgap}{\hspace{2pt}}
\begin{tabular}{@{}lcc *{3}{c@{\spkgap}c@{\spkgap}c} ccc@{}}
\toprule
\multirow{2}{*}{\textbf{Model}} &
\textbf{Latency} &
\multirow{2}{*}{\textbf{RTF}} &
\multicolumn{3}{c}{\textbf{NOTSOFAR}} &
\multicolumn{3}{c}{\textbf{VoxConverse}} &
\multicolumn{3}{c}{\textbf{DIHARD3}} &
\textbf{AMI} &
\textbf{MSDWild} &
\textbf{RAMC} \\
& (ms) & &
$\leq$4 spk & $\geq$5 spk & all &
$\leq$4 spk & $\geq$5 spk & all &
$\leq$4 spk & $\geq$5 spk & all &
3--4 spk & 2--4 spk & 2 spk \\
\midrule
\multicolumn{15}{l}{\textit{\textbf{Consider Overlap}}} \\
diart                & 1000 & 0.16 & 47.91 & 59.89 & 54.16 & 15.21 & 19.17 & 18.20 & 20.39 & 52.89 & 26.73 & 33.37 & 35.83 & 25.87 \\
LS-EEND              & 1000 & 0.09 & 37.57 & 53.08 & 45.67 & 23.01 & 44.66 & 39.36 & 15.94 & 44.02 & 21.41 & \textbf{21.81} & 31.28 & 32.98 \\
Streaming Sortformer & 1040 & 1.54 & \textbf{16.89} & \textbf{37.49} & \textbf{27.63} & \textbf{6.34} & 24.53 & 20.08 & \textbf{15.16} & \textbf{41.86} & \textbf{20.36} & 30.25 & \textbf{23.62} & 25.27 \\
FASTDIAR             & 960 &  0.19 & 34.59 & 44.25 & 39.63 & 13.76 & \textbf{15.49} & \textbf{15.07} & 29.42 & 45.62 & 32.57 & 33.36 & 32.61 & \textbf{21.61} \\
\midrule
\multicolumn{15}{l}{\textit{\textbf{Exclude overlap}}} \\
diart                & 1000 & 0.16 & 42.56 & 54.16 & 48.01 & 12.81 & 17.17 & 16.12 & 17.40 & 46.14 & 22.04 & 28.79 & 32.06 & 25.87 \\
LS-EEND              & 1000 & 0.09 & 34.93 & 49.10 & 41.58 & 22.43 & 44.41 & 39.07 & 14.28 & 36.19 & 17.82 & \textbf{15.98} & 30.20 & 32.98 \\
Streaming Sortformer & 1040 & 1.54 & \textbf{14.48} & 33.45 & 23.39 & \textbf{5.41} & 23.55 & 19.15 & \textbf{14.08} & 34.25 & \textbf{17.34} & 25.56 & \textbf{20.81} & 25.27 \\
FASTDIAR             & 960 & 0.19  & 19.23 & \textbf{24.25} & \textbf{21.58} & 10.70 & \textbf{12.88} & \textbf{12.35} & 24.73 & \textbf{32.95} & 26.06 & 22.15 & 24.69 & \textbf{21.61} \\
% \midrule
% Pyannote 3.1         & 8.11M & 2.26 & $\infty$ & - & - & 21.15 & - & - & 9.90 & - & - & 21.05 & 22.38 & 21.80 & 22.11  \\
% Pyannote community-1 & 8.11M & 2.57 & $\infty$ & - & - & 18.13 & - & - & 9.85 & - & - & 16.62 & 20.70 & 19.32 & 20.84 \\
% DiariZen-base        & 31.8M & 2.58 & $\infty$ & - & - & 13.87 & - & - & 8.71 & - & - & 12.54 & 14.59 & 13.46 & 11.37 \\
% DiariZen-large-v2    & 76.1M & 5.25 & $\infty$ & - & - & 12.76 & - & - & 8.20 & - & - & 11.28 & 12.94 & 13.73 & 10.99  \\
\bottomrule
\end{tabular}
\vspace{-0.6em}
\end{table*}

\subsection{Online clustering}
\label{subsec:clustering}

The algorithm is built around embedding confidence, the cosine similarity
$\langle \e_t, \e_{t-D} \rangle$ between the current embedding and the one $D$
frames back. High confidence means the embedding has been stable over the last
$D$ frames and the encoder has enough same-speaker context. Low confidence signals a speaker change,
silence or noise. Only confident frames update the cluster model, so centroids
are never degraded, while the rest are still labelled. Without it, an embedding
straddling a speaker change would drag both centroids together.

% later, no earlier and no later, so the delay is fixed rather than merely bounded

Algorithm~\ref{alg:clustering} summarizes the procedure. Every cluster $k$ keeps an
anchor set $A_k$ of confident frames, represented by the $\ell_2$-normalized
median $\m_k$ of their embeddings, which resists outliers better than the mean. A
confident frame joins the closest cluster when similarity exceeds
$\theta_{\text{clust}}$, and opens a new one otherwise. Labels are not emitted
immediately. Frame $t$ leaves the system when frame $t + M$ arrives, $M = 12$
frames or 960~ms fixed delay. The frames in between form a correction window.
What is emitted for frame $t$ is the cluster the stream settles on $M$ frames
later: when frame $t + M$ is confident, frame $t$ leaves with its label, decided
on 960~ms of context that frame $t$ alone does not carry, and an online merge can
revise that label on the way out, so a speaker change seen at $t + M$ sharpens
the boundary before it. When frame $t + M$ is unconfident, frame $t$ has run out
of context and joins the closest median at whatever similarity, without entering
an anchor set, so a guess never shapes a centroid. Without the
window a low-confidence stretch holds a frame back for as long as it lasts, and
the delay grows without bound exactly where fast turns, overlap and noise make it
longest. An off-the-shelf VAD \cite{silero2024vad} runs on the
same input stream, and every non-speech frame is emitted as silence instead of a
cluster label. Whenever a new cluster receives its first speech frame, an online merge links all
cluster pairs whose medians exceed $\theta_{\text{merge}}$ and fuses the connected
components, repairing over-segmentation. The
same merge can be optionally used over the whole recording at end of stream, as a
label correction. When the stream ends, the frames still inside the window are
emitted with their closest cluster. The assignment step follows incremental centroid
clustering as used in online linking \cite{mansfield2018links} and in diart
\cite{coria2021diart}, which map each embedding to the nearest centroid and open
a speaker when the distance exceeds a threshold. What we add is the confidence
gate. Confidence has been used in diarization before, but as a segment-level
score computed after clustering, from the similarity of a segment to its assigned
centroid, in order to flag likely errors for downstream consumers
\cite{chowdhury2024confidence}. Ours needs no centroid and no label, arrives with
the frame itself, and decides prospectively which frames may update a speaker
model. The three thresholds are tuned once on the development data of
Section~\ref{subsec:datasets} and then kept fixed for every experiment.

\begin{algorithm}[t]
\caption{Similarity-driven online clustering}
\vspace{0.25em}
\label{alg:clustering}
\begin{pseudocode}
\alghead{\kw{Input:} embeddings $\e_t$, VAD $v_t$, look-back $D=10$, emit delay $M=12$;
thresholds: $\theta_{\text{conf}}=0.8$, $\theta_{\text{clust}}=0.4$, $\theta_{\text{merge}}=0.8$}
\algline{$\mathcal{C} \gets \emptyset$, \; $P \gets \emptyset$}
\algline{\kw{for} $t = 0, 1, 2, \dots$ \kw{do}}
\algline[1em]{\kw{if} $t \geq D$ \kw{and} $\langle \e_t, \e_{t-D} \rangle \geq \theta_{\text{conf}}$ \kw{then}}
\algline[2em]{$k \gets \arg\max_{j \in \mathcal{C}} \langle \e_t, \m_j \rangle$}
\algline[2em]{\kw{if} $\mathcal{C} = \emptyset$ \kw{or} $\langle \e_t, \m_k \rangle < \theta_{\text{clust}}$ \kw{then}}
\algline[3em]{$k \gets \textsc{NewCluster}(\mathcal{C})$}
\algline[2em]{$A_k \gets A_k \cup \{t\}$, \; $\m_k \gets \textsc{Median}(\{\e_i\}_{i \in A_k})$}
\algline[2em]{$R \gets \{i \in P : i \leq t - M\}$, \; $P \gets (P \setminus R) \cup \{t\}$}
\algline[2em]{$\ell_i \gets k$ \kw{for} $i \in R$ \algcom{retroactive labels}}
\algline[2em]{\kw{if} $k$ is new \kw{and} $R$ contains speech \kw{then} $\textsc{Merge}(\mathcal{C}, \theta_{\text{merge}})$}
\algline[1em]{\kw{else} $P \gets P \cup \{t\}$ \algcom{unconfident, defer}}
\algline[1em]{\kw{for} $i \in P$ \kw{with} $i \leq t - M$ \kw{do} \algcom{out of context}}
\algline[2em]{$P \gets P \setminus \{i\}$, \; $\ell_i \gets \arg\max_{j \in \mathcal{C}} \langle \e_i, \m_j \rangle$ \algcom{no anchor}}
\algline[1em]{\kw{emit} $(t-M, \ell_{t-M})$ \kw{if} $v_{t-M} = 1$ \kw{else} $(t-M, \text{silence})$}
\algline{assign the frames left in $P$ to their closest cluster}
\end{pseudocode}
\vspace{-0.6em}
\end{algorithm}

\section{Experiments}
\label{sec:experiments}

\subsection{Datasets}
\label{subsec:datasets}

\noindent\textbf{Training.} VoxCeleb2 \cite{chung2018voxceleb2} contains one
speaker per utterance, so multi-speaker examples are simulated on the fly. With
probability $0.75$ we draw crops $x_1$ and $x_2$ from two different speakers,
normalize both to unit RMS, and arrange them over $T = 8$~s in one of two
layouts. The first has one speaker change at $T/2$ preceded by an overlap of
$o$~ms, $\log o \sim \mathcal{N}(\mu, \sigma^2)$ truncated to $[150, 4000]$~ms and
excluded from the loss. The second alternates $x_1 \rightarrow x_2 \rightarrow x_1$
with segments of at least one second. The interferer is scaled to a signal-to-interference ratio
from $\mathcal{U}(-5, 5)$~dB, and the mixture is RMS-normalized so level carries
no cue. We augment with MUSAN \cite{snyder2015musan} and simulated
room impulse responses \cite{ko2017rirs} through WeSpeaker
\cite{wang2023wespeaker}. We further use the
medium split of LibriHeavyMix \cite{jin2024libriheavymix}, keeping unaugmented mixtures of two or
three speakers in which one covers 20--90\% of the audio. VoxCeleb2 is sampled 75\% of the time and LibriHeavyMix 25\%, with the
VoxConverse and NOTSOFAR development sets used to select the checkpoint and tune
the three clustering thresholds.

\noindent\textbf{Benchmarks.} We assess diarization on the evaluation sets of
NOTSOFAR \cite{vinnikov2024notsofar} (single channel), VoxConverse v0.3
\cite{chung2020voxconverse}, DIHARD3 \cite{ryant2021dihard3}, AMI \cite{carletta2005ami} (IHM-MIX), MSDWild \cite{liu2022msdwild} (few), and RAMC \cite{yang2022ramc}. They cover multiple
languages, various distances and 2 to 21 speakers per recording.
Streaming Sortformer is capped at four speakers, so for the three corpora
crossing that boundary, we also report DER split at it.
Speaker recognition is measured on
VoxCeleb1-E \cite{nagrani2017voxceleb}.

\subsection{Model and training}
\label{subsec:model}

We distil an utterance-level teacher into a frame-level student, as in
\cite{cordlandwehr2023frame}, but maximize cosine similarity \cite{jung2019short}
rather than minimize squared error. The teacher embeds the clean, unaugmented crop of speaker $s$ as $\z_s$, and the student embeds the augmented mixture frame
by frame. The objective is%
% let the lead-in finish in the left column instead of carrying a line
% across to the right one
\displaywidowpenalty=0 \predisplaypenalty=0
\begin{equation}
  \L = 1 - \frac{1}{|F|} \sum_{t \in F} \langle \e_t, \z_{s(t)} \rangle ,
\end{equation}
where $s(t)$ is the speaker active at frame $t$ and $F$ holds the frames with
exactly one active speaker, so overlap is ignored.

The teacher is ReDimNet2-B6 trained on VoxBlink2 \cite{lin2024voxblink2} and
VoxCeleb2. Students are B2 (Small), B4 (Medium), and B6 (Large), differing only in
width and FLOPs. They reuse the 2D convolution
weights of the matching ReDimNet2, initialize fully connected layers from
scratch, and update every weight. Training runs on one RTX 6000 Ada GPU in bf16, on 8~s
segments, batch 20, 100 epochs of 2000 steps, gradients clipped at 0.5, and AdamW \cite{loshchilov2019adamw}
warmed over one epoch to a peak of $1 \times 10^{-3}$ then decayed linearly.

\subsection{Baseline models}
\label{subsec:baselines}

We compare against open-source streaming systems: Streaming Sortformer
\cite{medennikov25_interspeech} v2.1 at 1040~ms latency, diart
\cite{coria2021diart} at 1000~ms with recommended thresholds for DIHARD3, AMI and VoxConverse, and LS-EEND
\cite{liang2025lseend} at 1000~ms. All are reproduced with released code and checkpoints. LS-EEND ships
one model per corpus: we use its DIHARD3 checkpoint by default, the strongest on
our development data, and its AMI checkpoint for that test, so it is in-domain on
DIHARD3 and AMI and out-of-domain elsewhere.
To isolate the encoder we build ReDimNet2-B6-S, the unmodified
ReDimNet2-B6 applied to two-second windows advanced by 80~ms,
reproducing our receptive field and embedding rate and feeding the same VAD and
clustering. 

\subsection{Evaluation protocol}
\label{subsec:evaluation}

We report diarization error rate (DER) with a collar of 0, with and without
overlap, and split it by the number of speakers. Latency is the delay between an audio frame entering the system and its label
leaving it, fixed at $M = 12$ frames, that is 960~ms, for every frame of every
recording. It covers the $D = 10$ frame look-back the system fills
before it can emit at all. The real-time factor (RTF) is processing time over input duration on one thread of an AMD Threadripper PRO 5995WX CPU, covering the encoder, the VAD and
the clustering for FASTDIAR, and we report equal error rate (EER) for verification.

 % so 960~ms is both the worst case and the mean

\section{Results and Discussion}
\label{sec:results}

\begin{table}[t]
\centering
\caption{VoxCeleb1-E cleaned protocol. Context and shift in ms. For the
frame-level models a pair score averages over all embedding pairs. X-EER embeds
the enrollment side of each trial with ReDimNet2-B6.}
\vspace{0.25em}
\label{tab:vox1}
\small
\setlength{\tabcolsep}{3pt}
\begin{tabular}{lccccc}
\toprule
\textbf{Model} & \textbf{Context} & \textbf{Shift} & \textbf{EER(\%)} & \textbf{X-EER(\%)} & \textbf{RTF} \\
\midrule
ReDimNet2-B6     & $\infty$ & --  & \textbf{0.34} & --            & --   \\
ReDimNet2-B6-S   & 2000     & 80  & 0.44          & \textbf{0.39} & 4.60 \\
Proposed (Large) & 2000     & 80  & 1.51          & 0.86          & \textbf{0.17} \\
\bottomrule
\end{tabular}
\vspace{-0.6em}
\end{table}

\noindent\textbf{Diarization accuracy.} Results in Table~\ref{tab:diarization}
vary greatly under each protocol. Without overlap, FASTDIAR leads on NOTSOFAR,
VoxConverse and RAMC, and trails only the in-domain LS-EEND on AMI. With overlap counted, Streaming Sortformer moves ahead of it on
NOTSOFAR, DIHARD3, AMI and MSDWild, and FASTDIAR keeps VoxConverse and RAMC, the
corpora with little or no overlap.

\noindent\textbf{Speaker count.} Streaming Sortformer is built around a cache of
at most four speakers and degrades sharply beyond it: excluding overlap, its DER grows by a factor of 2.3 to 4.4 across the
four-speaker boundary, against 1.2 to 1.5 for FASTDIAR, which leads the $\geq$5
column on NOTSOFAR and VoxConverse. Nothing in the encoder or the clustering is
tied to a speaker count, which a streaming system rarely knows.

\noindent\textbf{Baselines and efficiency.} LS-EEND leads on DIHARD3 and AMI,
its two in-domain corpora, and falls behind on the other four, while FASTDIAR
beats diart almost everywhere. On one CPU thread FASTDIAR, diart and LS-EEND all
run several times faster than real time, while Streaming Sortformer needs a GPU.
FASTDIAR has the lowest algorithm latency.

\noindent\textbf{Generalization.} FASTDIAR trains on simulated VoxCeleb2
mixtures and LibriHeavyMix, with no real diarization corpus. Tuning the checkpoint and thresholds on their development sets leaves VoxConverse and NOTSOFAR not fully unseen, while the other four run untouched. The price shows on DIHARD3, where FASTDIAR is last under both
protocols. Our simulated mixtures cover only a fraction of the scenarios it targets, and every system ahead of it has seen DIHARD3: LS-EEND trains on
it, Sortformer on a mixture including its development set, and diart tunes
its thresholds there.

\noindent\textbf{Encoder trade-off.} As Table~\ref{tab:vox1} shows, the chunked
model keeps verification accuracy but re-encodes every sample many times,
which rules out streaming on CPU. The proposed encoder is more than $25$ times
faster, but it cannot pool over an utterance, so its EER is around 3 times
worse. Distillation leaves the student in the teacher's space, so the two are directly comparable. Embedding one side of every trial with the offline teacher
(X-EER) cuts the proposed encoder from 1.51 to 0.86 and the chunked model from
0.44 to 0.39, so one pooled side recovers more than half of the gap to the teacher. Enrollment is rarely time-critical, so a target-speaker profile, or
the first centroid of a new cluster, can be built offline and then matched
against streaming embeddings.

\begin{table}[t]
\centering
\caption{Ablation on model scaling and offline post-processing. All models
exclude overlapped speech and have 960~ms latency.}
\vspace{0.25em}
\label{tab:ablation}
\small
\setlength{\tabcolsep}{4pt}
\begin{tabular}{lcccc}
\toprule
\textbf{Model} &
\textbf{\#Params} &
\textbf{RTF} &
\textbf{VoxConverse} &
\textbf{RAMC}
\\
\midrule
ReDimNet2-B6-S   & 12.3M & 4.85 & 14.79 & 27.97 \\
\midrule
Small            & 2.7M & 0.05 & 16.49 & 31.59 \\
Medium           & 4.6M & 0.09 & 14.06 & 24.21 \\
Large (FASTDIAR) & \multirow{2}{*}{10.4M} & \multirow{2}{*}{0.19} & 12.35 & 21.61 \\
\hspace{0.2cm}+ post-processing & & & \textbf{11.77} & \textbf{20.08} \\
\bottomrule
\end{tabular}
\vspace{-0.6em}
\end{table}

% \textbf{NOTSOFAR} &
% \textbf{VoxConverse}
% ReDimNet2-B6-S   & 12.3M & 4.85 & 22.34 & 14.10 \\
% \midrule
% Small            & 2.7M & 0.05 & 29.79 & 15.59 \\
% Medium           & 4.6M & 0.09 & 21.81 & 12.85 \\
% Large (FASTDIAR) & \multirow{2}{*}{10.4M} & \multirow{2}{*}{0.19} & 19.36 & 11.77 \\
% \hspace{0.2cm}+ post-processing & & & \textbf{19.34} & \textbf{10.79} \\

\noindent\textbf{Ablation.} Table~\ref{tab:ablation} answers two questions. Does
the frame-level encoder have to be trained, or would chunking an off-the-shelf
model do? ReDimNet2-B6-S shares the receptive field and embedding rate and is the
better verification model, yet Medium and Large beat it on diarization while being smaller and faster: a stronger verification model is not automatically a better
diarizer. Is the result specific to one size? Accuracy improves monotonically
from Small to Large, mirroring the scaling of ReDimNet2, so the recipe
extrapolates rather than working by accident. Post-processing merges
similar clusters at the end of the stream, outside the streaming path, showing
how far offline revision can go.

\noindent\textbf{Limitations.} FASTDIAR's main limitation is overlapped speech: it assigns one label per frame, so every overlapped region counts as a miss. This explains most of the gap between the halves of
Table~\ref{tab:diarization}, and it comes from the decoding stage, not the
encoder, which is trained on frames with one active speaker. In future work we
plan to keep overlapped frames in the objective and predict several labels per frame, changing neither the encoder nor its inference cost.

\section{Conclusion}
\label{sec:conclusion}

We presented FASTDIAR, a streaming diarization system built on a causal
frame-level speaker encoder that reads the stream once and emits an embedding
every 80~ms from a two-second past window. Removing temporal pooling costs verification accuracy but buys sharp
speaker-change localization and five times real-time speed on one CPU thread. With confidence-gated online clustering at a fixed 960~ms delay, it is the most
accurate streaming diarizer on low-overlap benchmarks, degrades far less than
cache-based systems beyond four speakers, and needs no diarization corpus in training. The encoder is the reusable
part: any real-time model needing frame-level
speaker identity can read the same stream, and the diarizer here is only one of its consumers.

\bibliographystyle{IEEEbib}
\bibliography{refs}

@inproceedings{desplanques2020ecapa,
  title     = {{ECAPA-TDNN}: Emphasized Channel Attention, Propagation and Aggregation in {TDNN} Based Speaker Verification},
  author    = {Brecht Desplanques and Jenthe Thienpondt and Kris Demuynck},
  booktitle = {Proc. Interspeech},
  year      = {2020},
}

@inproceedings{yakovlev2024redimnet,
  title     = {Reshape {D}imensions {N}etwork for {S}peaker {R}ecognition},
  author    = {Ivan Yakovlev and others},
  booktitle = {Proc. Interspeech},
  year      = {2024},
}

@unpublished{yakovlev2026redimnet2,
  title     = {{ReDimNet2}: {S}caling {S}peaker {V}erification via {T}ime-{P}ooled {D}imension {R}eshaping},
  author    = {Ivan Yakovlev and Anton Okhotnikov},
  note      = {to appear in \emph{Proc. Interspeech}},
  year      = {2026},
}

@inproceedings{Bredin23,
  author={Hervé Bredin},
  title={{pyannote.audio 2.1 speaker diarization pipeline: principle, benchmark, and recipe}},
  year=2023,
  booktitle={Proc. Interspeech},
}

@inproceedings{han2025leveraging,
  title={Leveraging self-supervised learning for speaker diarization},
  author={Han, Jiangyu and others},
  booktitle={Proc. ICASSP},
  year={2025}
}

@inproceedings{fujita2019eend,
  title     = {End-to-End Neural Speaker Diarization with Self-Attention},
  author    = {Yusuke Fujita and Naoyuki Kanda and others},
  booktitle = {Proc. ASRU},
  year      = {2019},
}

@inproceedings{horiguchi2020eendeda,
  title     = {End-to-End Speaker Diarization for an Unknown Number of Speakers with Encoder-Decoder Based Attractors},
  author    = {Shota Horiguchi and others},
  booktitle = {Proc. Interspeech},
  year      = {2020},
}

@article{park2024sortformer,
  title   = {Sortformer: Seamless Integration of Speaker Diarization and {ASR} by Bridging Timestamps and Tokens},
  author  = {Taejin Park and Ivan Medennikov and others},
  journal = {arXiv:2409.06656},
  year    = {2024},
}

@inproceedings{medennikov25_interspeech,
  title     = {{Streaming Sortformer: Speaker Cache-Based Online Speaker Diarization with Arrival-Time Ordering}},
  author    = {Ivan Medennikov and Taejin Park and others},
  year      = {2025},
  booktitle = {{Proc. Interspeech}},
}

@inproceedings{liang2024fseend,
  title     = {Frame-Wise Streaming End-to-End Speaker Diarization with Non-Autoregressive Self-Attention-Based Attractors},
  author    = {Di Liang and Nian Shao and Xiaofei Li},
  booktitle = {Proc. ICASSP},
  year      = {2024},
}

@ARTICLE{liang2025lseend,
  author={Liang, Di and others},
  journal={IEEE Trans. Audio, Speech, Lang. Process.}, 
  title={{LS-EEND}: Long-Form Streaming End-to-End Neural Diarization With Online Attractor Extraction}, 
  year={2025},
}

@inproceedings{xue2021stb,
  title     = {Online End-to-End Neural Diarization with Speaker-Tracing Buffer},
  author    = {Yawen Xue and others},
  booktitle = {Proc. SLT},
  year      = {2021},
}

@inproceedings{han2021bweda,
  title     = {{BW-EDA-EEND}: Streaming End-to-End Neural Speaker Diarization for a Variable Number of Speakers},
  author    = {Eunjung Han and Chul Lee and Andreas Stolcke},
  booktitle = {Proc. ICASSP},
  year      = {2021},
}

@article{horiguchi2023gla,
  title   = {Online Neural Diarization of Unlimited Numbers of Speakers Using Global and Local Attractors},
  author  = {Shota Horiguchi and Shinji Watanabe and Paola Garcia and Yuki Takashima and Yohei Kawaguchi},
  journal = {IEEE/ACM Trans. Audio, Speech, Lang. Process.},
  year    = {2023},
}

@inproceedings{coria2021diart,
  title     = {Overlap-Aware Low-Latency Online Speaker Diarization Based on End-to-End Local Segmentation},
  author    = {Juan M. Coria and Herv{\'e} Bredin and Sahar Ghannay and others},
  booktitle = {Proc. ASRU},
  year      = {2021},
}

@inproceedings{cordlandwehr2023frame,
  title     = {Frame-Wise and Overlap-Robust Speaker Embeddings for Meeting Diarization},
  author    = {Tobias Cord-Landwehr and Christoph Boeddeker and others},
  booktitle = {Proc. ICASSP},
  year      = {2023},
}

@inproceedings{chowdhury2024confidence,
  title     = {Investigating Confidence Estimation Measures for Speaker Diarization},
  author    = {Anurag Chowdhury and Abhinav Misra and others},
  booktitle = {Proc. Interspeech},
  year      = {2024},
}

@inproceedings{zhang2019uisrnn,
  title     = {Fully Supervised Speaker Diarization},
  author    = {Aonan Zhang and Quan Wang and Zhenyao Zhu and John Paisley and others},
  booktitle = {Proc. ICASSP},
  year      = {2019},
}

@inproceedings{xia2022turn,
  title     = {Turn-to-Diarize: Online Speaker Diarization Constrained by Transformer Transducer Speaker Turn Detection},
  author    = {Wei Xia and Han Lu and Quan Wang and others},
  booktitle = {Proc. ICASSP},
  year      = {2022},
}

@article{mansfield2018links,
  title   = {Links: A High-Dimensional Online Clustering Method},
  author  = {Philip Andrew Mansfield and others},
  journal = {arXiv:1801.10123},
  year    = {2018},
}

@inproceedings{jung2019short,
  title     = {Short Utterance Compensation in Speaker Verification via Cosine-Based Teacher-Student Learning of Speaker Embeddings},
  author    = {Jee Weon Jung and others},
  booktitle = {Proc. ASRU},
  year      = {2019},
}

@inproceedings{chung2018voxceleb2,
  title     = {{VoxCeleb2}: Deep Speaker Recognition},
  author    = {Joon Son Chung and Arsha Nagrani and others},
  booktitle = {Proc. Interspeech},
  year      = {2018},
}

@inproceedings{nagrani2017voxceleb,
  title     = {{VoxCeleb}: A Large-Scale Speaker Identification Dataset},
  author    = {Arsha Nagrani and others},
  booktitle = {Proc. Interspeech},
  year      = {2017},
}

@inproceedings{lin2024voxblink2,
  title     = {{VoxBlink2}: A 100{K}+ Speaker Recognition Corpus and the Open-Set Speaker-Identification Benchmark},
  author    = {Yuke Lin and Ming Cheng and others},
  booktitle = {Proc. Interspeech},
  year      = {2024},
}

@article{snyder2015musan,
  title   = {{MUSAN}: A Music, Speech, and Noise Corpus},
  author  = {David Snyder and Guoguo Chen and Daniel Povey},
  journal = {arXiv:1510.08484},
  year    = {2015},
}

@inproceedings{ko2017rirs,
  title     = {A Study on Data Augmentation of Reverberant Speech for Robust Speech Recognition},
  author    = {Tom Ko and Vijayaditya Peddinti and Daniel Povey and Michael L Seltzer and others},
  booktitle = {Proc. ICASSP},
  year      = {2017},
}

@inproceedings{wang2023wespeaker,
  title     = {We{S}peaker: A Research and Production Oriented Speaker Embedding Learning Toolkit},
  author    = {Hongji Wang and Chengdong Liang and Shuai Wang and Zhengyang Chen and others},
  booktitle = {Proc. ICASSP},
  year      = {2023},
}

@inproceedings{jin2024libriheavymix,
  title     = {{LibriheavyMix}: A 20,000-Hour Dataset for Single-Channel Reverberant Multi-Talker Speech Separation, {ASR} and Speaker Diarization},
  author    = {Zengrui Jin and Yifan Yang and Mohan Shi and Wei Kang and  Xiaoyu Yang and others},
  booktitle = {Proc. Interspeech},
  year      = {2024},
}

@inproceedings{vinnikov2024notsofar,
  title     = {{NOTSOFAR-1} Challenge: New Datasets, Baseline, and Tasks for Distant Meeting Transcription},
  author    = {Alon Vinnikov and Amir Ivry and Aviv Hurvitz and others},
  booktitle = {Proc. Interspeech},
  year      = {2024},
}

@inproceedings{chung2020voxconverse,
  title     = {Spot the Conversation: Speaker Diarisation in the Wild},
  author    = {Joon Son Chung and others},
  booktitle = {Proc. Interspeech},
  year      = {2020},
}

@inproceedings{ryant2021dihard3,
  title     = {The Third {DIHARD} Diarization Challenge},
  author    = {Neville Ryant and Prachi Singh and others},
  booktitle = {Proc. Interspeech},
  year      = {2021},
}

@inproceedings{liu2022msdwild,
  title     = {{MSDWild}: Multi-Modal Speaker Diarization Dataset in the Wild},
  author    = {Tao Liu and others},
  booktitle = {Proc. Interspeech},
  year      = {2022},
}

@inproceedings{yang2022ramc,
  title     = {Open Source {MagicData-RAMC}: A Rich Annotated {M}andarin Conversational Speech Dataset},
  author    = {Zehui Yang and Yifan Chen and Lei Luo and others},
  booktitle = {Proc. Interspeech},
  year      = {2022},
}

@inproceedings{carletta2005ami,
  title     = {The {AMI} Meeting Corpus: A Pre-Announcement},
  author    = {Jean Carletta and Simone Ashby and others},
  booktitle = {Proc. Machine Learning for Multimodal Interaction (MLMI)},
  year      = {2005},
}

@misc{silero2024vad,
  title        = {Silero {VAD}: Pre-Trained Enterprise-Grade Voice Activity Detector},
  author       = {{Silero Team}},
  howpublished = {\url{https://github.com/snakers4/silero-vad}},
  year         = {2024},
}

@inproceedings{loshchilov2019adamw,
  title     = {Decoupled Weight Decay Regularization},
  author    = {Ilya Loshchilov and Frank Hutter},
  booktitle = {Proc. ICLR},
  year      = {2019},
}

\end{document}